\documentclass{aastex6}

\usepackage{graphicx}
\usepackage{hyperref}
\usepackage{ulem}
\usepackage{txfonts}
\usepackage{color}
\usepackage{multirow}
\usepackage{rotating}

\newcommand{\be}{\begin{equation}}
\newcommand{\ee}{\end{equation}}
\newcommand{\beq}{\begin{eqnarray}}
\newcommand{\eeq}{\end{eqnarray}}

\def\nue{\mathrel{{\nu_e}}}

\def\nux{\mathrel{{\nu_x}}}

\def\barnue{\mathrel{{\bar \nu}_e}}

\def \lta {\mathrel{\vcenter{\hbox{$<$}\nointerlineskip\hbox{$\sim$}}}}
\def \gta {\mathrel{\vcenter{\hbox{$>$}\nointerlineskip\hbox{$\sim$}}}}

\def\t13{\mathrel{{\theta_{13}}}}
\def\y12{\mathrel{{\tan^2 \theta_{12}}}}
\def\c2{\mathrel{{\chi^2 }}}

\def\msun{\mathrel{{M_{\odot} }}}

\def\qth{\mathrel{{Q_{th} }}}

\newcommand{\n}{neutrino}
\newcommand{\ns}{neutrinos}
\newcommand{\ps}{presupernova}
\newcommand{\sn}{supernova}
\newcommand{\sne}{supernovae}
\newcommand{\qv}{$Q$-value}

\newcommand\mesh{\ensuremath{\delta_{\rm{mesh}}}} 
\newcommand{\maxdm}{{\ensuremath{\Delta {\mathrm M}_{\rm{max}}}}} 
\newcommand{\net}{{\tt mesa\_{204}.net}}

\begin{document}

\title{Presupernova neutrinos: realistic emissivities from stellar evolution}

\author{Kelly M.\ Patton}
\affiliation{Department of Physics, Arizona State University, Tempe, AZ 85287-1504 USA \\
Institute for Nuclear Theory, University of Washington, Seattle, WA 98195 USA}

\author{Cecilia Lunardini}
\affiliation{Department of Physics, Arizona State University, Tempe, AZ 85287-1504 USA}

\author{Robert J.\ Farmer}
\affiliation{School of Earth and Space Exploration, Arizona State University, Tempe, AZ 85287-1504 USA}

\email{kmpatton@uw.edu}
\email{cecilia.lunardini@asu.edu}
\email{rjfarmer@asu.edu}

\begin{abstract}
  We present a new calculation of \n\ emissivities and energy spectra from a massive star going through the advanced stages of nuclear burning (presupernova) in the months before becoming a supernova.  The contributions from beta decay and electron capture, pair annihilation, plasmon decay, and the photoneutrino process are modeled in detail, using updated tabulated nuclear rates. We also use realistic conditions of temperature, density, electron fraction and nuclear isotopic composition of the star from the state of the art stellar evolution code MESA. Results are presented for a set of progenitor stars with mass between 15 $M_\odot$ and 30 $M_\odot$.  It is found that beta processes contribute substantially to the \n\ emissivity above realistic detection thresholds of few MeV, at selected positions and  times in the evolution of the star.  
\end{abstract}

\date{\today}

\maketitle


\section{Introduction}
\label{sec:intro}

A very luminous burst of \ns\ is the first signal that we receive -- possibly together with gravitational waves --   informing us that a star's core has collapsed, and that just a few hours afterwards the initially masked electromagnetic components will escape, becoming a \sn.  Since their first (and only) detection in 1987 \citep{Hirata:1987, Bionta:1987, Alekseev:1987}, \ns\ from stellar collapse have been studied in their rich phenomenology, including their role in the dynamics of collapse and explosion, the effects on nucleosynthesis processes in the stellar matter, and the complicated pattern of \n\ flavor oscillations inside the star and in the Earth (see e.g. \cite{Mirizzi:2015eza}).

Interestingly, \ns\ can also offer a unique signature of the stages of stellar evolution that lead up to collapse (``presupernova"), on which still little is known, at least observationally, compared to the dramatic post-collapse events.   As a massive star ($M \gta 8 M_\odot$, with $M_\odot$ the mass of the Sun) nears the end of its lifetime, the chain of nuclear burning in its core and inner shells proceeds through the fusion of progressively heavier elements.  The central temperature and density of the star increase dramatically, resulting in an equally dramatic increase in the flux and average energies of neutrinos emitted. These \ns\ could become detectable months before the collapse, during the oxygen-burning phase, for the closest supernova candidate, Betelgeuse. Days or hours before (silicon-burning phase) might be more realistic for a star a few kiloparsecs away \citep{Odrzywolek:2010zt,Asakura:2015sw}.  

Although challenging, the observation of \ps\ \ns\ would be extremely rewarding: it would offer an unprecedented direct probe of nuclear fusion beyond hydrogen and helium, and give us a first-hand narrative of the very late stages of stellar evolution in terms of density and temperature near the star's core.  Considering the excellent timing resolution of current \n\ detectors, this narrative could be seen in real time, and would be a precious alert of the upcoming post-collapse \n\ burst and supernova. 

Besides their detection, the production and propagation of \ps\ \ns\ are important ingredients of models of stellar evolution.  In the later stages of nuclear burning, neutrinos become the main source of energy loss while also increasing the entropy of the star as it nears core collapse \citep{Woosley:2002qb}.  The physics of these \ns\ is interesting also as an important application of the more general problem of neutrino emission in hot and dense stellar matter. 

With these motivations, studies have been conducted on the \n\ emissivity of stars in the \ps\ stage. 
Most of the literature so far has focused on \ns\ produced via thermal processes,  for representative conditions (temperature, density and chemical potential) of the stellar matter.  The earliest works \citep{Odrzywolek:2003ij, Odrzywolek:2004rm, Kutschera:2009ff} included only the pair annihilation process, and parameters typical of the Si burning phase.  The possibility to detect the resulting \n\ flux was discussed, with encouraging conclusions. A more detailed study of \ps\ \ns\ from thermal processes, and their detectability, has appeared recently \citep{Kato:2015ec}, including pair annihilation and plasmon decay.  Rather than representative parameters, this work uses realistic, time-evolving profiles of temperature, density and chemical potential from numerical models of stellar evolution \citep{Takahashi:2013ena}.  A second paper by a subset of the authors of \citet{Kato:2015ec} explores in detail the pair annihilation neutrino spectra and detection potential in both current and future detectors, with emphasis on what the variation in the neutrino signal can indicate about stellar evolution \citep{2016PhRvD..93l3012Y}. 

Until now, the role of $\beta$ processes in \ps\ \ns\ has been discussed only in the basics, in the works of Odrzywolek and Heger \citep{Odrzywolek:2009yl,Odrzywolek:2010zt}. There, arguments of nuclear statistical equilibrium or $\alpha$-networks are used to determine isotopic composition. In \cite{Odrzywolek:2010zt}, it is explicitly emphasized that both methods are inadequate, and that a full, self-consistent stellar evolution simulation, with a large and accurate nuclear reaction network is ultimately needed.

In this work, such rigorous approach is realized for the first time. We present a new, comprehensive calculation of the \ps\ \n\ emission, which includes, in addition to the main thermal processes (pair annihilation, plasmon decay, and the photoneutrino process), a detailed treatment of   $\beta$ decay and electron capture. These processes are modeled using updated nuclear rate tables \citep{Langanke:2001, Oda:1994} as a supplement to  the classic ones by Fuller, Fowler and Newman \citep{Fuller:1980,Fuller:1982a, Fuller:1982, Fuller:1985}. The relevant microphysics is then applied to a realistic star using the detailed,  time-evolving profiles of temperature, density, and nuclear isotopic composition from  the state-of-the-art stellar evolution code MESA (Modules for Experiments in Stellar Astrophysics) \citep{Paxton:2010ig, Paxton:2013jy, Paxton:2015vl}. We place emphasis on modeling of the \n\ spectrum above a realistic detection threshold of 2 MeV; this requires including certain $\beta$ processes that are subdominant in the total energy budget of the star.

The paper is structured as follows. After a summary of background information (sec. \ref{sec:production}), the relevant formalism of \n\ emissivities and spectra is discussed  in sec. \ref{sec:beta} for $\beta$-processes, and in sec. \ref{sec:thermal} for thermal processes.  In sec. \ref{sec:results} numerical results are shown for several steps of a star's \ps\ evolution, and for different progenitor stars, as modeled by MESA.  A discussion and final considerations are given in sec. \ref{sec:discussion}. 

\begin{deluxetable}{ccccc}
\tablecaption{Summary of the processes included in this work, with the main references to prior literature. \label{tab:microphys}}
\tablehead{ & \colhead{Processes} & & \colhead{Formulae} & \colhead{Main References}}
\startdata
 & & $\beta^{\pm}$ decay & $A(N,Z) \rightarrow A(N-1,Z+1) + e^{-} + \overline{\nu}_e $ & \\
  & & & $A(N,Z) \rightarrow A(N+1,Z-1) + e^{+} + \nu_e $ & \cite{Fuller:1980, Fuller:1982, Fuller:1982a, Fuller:1985},\\
  Beta  & & &  &  \cite{Langanke:2001},  \\
   & & $e^+$/$e^-$ capture & $A(N,Z) + e^{-} \rightarrow A(N+1,Z-1) + \nu_e $ & \cite{Oda:1994, Odrzywolek:2009yl} \\ \vspace{.2in}
  & & & $A(N,Z) + e^{+} \rightarrow A(N-1,Z+1) + \overline{\nu}_e$ & \\
  & & plasma & $\gamma^\ast \rightarrow \nu_\alpha  + \overline{\nu}_\alpha$  & \cite{Ratkovic:2003pu, Odrzywolek:2007kv}\\
  Thermal &  & photoneutrino & $e^{\pm} + \gamma \rightarrow e^{\pm} + \nu_\alpha + \overline{\nu}_\alpha $  & \cite{Dutta:2004ug} \\
  & & pair  & $e^+ + e^-  \rightarrow \nu_\alpha  + \overline{\nu}_\alpha$ & \cite{Misiaszek:2006jo} \\
\enddata
\end{deluxetable}%


\section{Neutrino production in a presupernova environment}
\label{sec:production}

About $\sim 10^3$ years before becoming a \sn, a star begins to experience the fusion of heavy (beyond helium) elements. First, carbon fusion is ignited; as the temperature and density increases, then the fusion of Ne, O, and Si take place in the core of the star.  Each stage is faster than the previous one:  the core O burning phase only lasts a few months, and the core Si burning only takes a few days \citep{Woosley:2002qb}.   Immediately before collapse, the star is characterized by a shell structure, with an iron core at the center, surrounded by shells where heavy element fusion is still taking place efficiently.  

In the increasingly dense and hot environment of a \ps, \ns\ are produced more and more abundantly through several processes, which we broadly categorize as $\beta$ and thermal. Here the \n\ emissivities and spectra are calculated for the four main processes, using analytic and semi-analytic results from the literature, as summarized in Table \ref{tab:microphys}.   Each process is discussed in detail in the subsections below. 

\subsection{$\beta$ Processes}
\label{sec:beta}

When a star reaches the presupernova phase, its nuclear isotopic composition is complex and constantly changing. Therefore, to calculate the $\nue$ and $\barnue$ fluxes from beta processes (Table \ref{tab:microphys}) requires information on a vast array of nuclear transitions. 

Here we use the rate tables compiled by Fuller, Fowler and Newman (FFN) \citep{Fuller:1980, Fuller:1982, Fuller:1982a, Fuller:1985}, Oda \textit{et al.} (OEA) \citep{Oda:1994}, and Langanke and Martinez-Pinedo (LMP) \citep{Langanke:2001}.  Each table uses shell model calculations, including experimental data when available, to find rates of electron (positron) capture and $\beta^{\pm}$ decays for a grid of temperature and density values, under the assumption that there is a strong contribution from Gamow-Teller (GT) transitions.  The FFN table covers isotopes with $21 \leq A \leq 60$, while OEA covers $17 \leq A \leq 39$ and LMP is calculated for $45 \leq A \leq 65$.  Where overlap between different tables occurs, precedence is given, in order, to  LMP, then OEA  and finally FFN. This is the same convention used by MESA \citep{Paxton:2010ig}.

Let us summarize the calculation of \n\ $\beta$ emissivities and spectra. The rate of weak decay from the $i^{th}$ parent state to the $j^{th}$ daughter state is written as \citep{Fuller:1980}
\begin{equation}
\lambda_{ij} = \log{2}\frac{f_{ij}(T,\rho,\mu_{e})}{\langle ft \rangle_{ij}}.
\label{eq:weakDecayRate}
\end{equation}
The quantity $\langle ft \rangle_{ij}$ is the comparative half-life for the process, and is related to the weak interaction matrix element.  For the tables of FFN, OEA, and LMP, the value of $\langle ft \rangle$ is taken either from experimental measurements or from estimates of the strength of Gamow-Teller and Fermi transitions.  

The function $f_{ij}(T, \rho, \mu_{e})$ is the phase space integral for the process.  The phase space of an outgoing electron with momentum $p'$ is given by 
 \begin{equation}
dn_{p'} = p'^{2} dp \left(1 - \frac{1}{1 + \exp{((E_{e'} - \mu_{e})/kT)}}\right),
\label{pout}
\end{equation} 
while for an incoming electron with momentum $p$ it is
 \begin{equation}
dn_{p} = p^{2} dp \left(\frac{1}{1 + \exp{((E_{e} - \mu_{e})/kT)}}\right).
\label{pin}
\end{equation} 
Here $p$ and  $E_{e} = \sqrt{p^{2} + m_{e}^{2}}$ are the momentum and energy of the electron, $\mu_{e}$ is the chemical potential, and $T$ is the temperature.  As in \citet{Odrzywolek:2009yl}, we define the chemical potential including the rest mass, so that $\mu_{e^{-}} = - \mu_{e^{+}}$.  The outgoing neutrinos are assumed to have no inhibition of the final state \citep{Fuller:1980}, so the phase space factor is simply $E_{\nu}^{2} dE_{\nu}$. In other words, while the incoming and outgoing electrons must conform to a Fermi-Dirac distribution, neutrinos have no such restriction.  

For a given nuclear transition, the $Q$-value is defined as \citep{Fuller:1980}
\begin{equation}
Q_{ij} = M_{p} - M_{d} + E_{i} - E_{j},
\label{eq:Qvalue}
\end{equation}
where $M_{p}$ and $E_i$ ($M_d$ and $E_j$) are  the mass and excitation energy of of the parent (daughter) nucleus.  Since we are interested in the rate as a function of neutrino energy, we can rewrite the electron phase space integrals, Eqs. (\ref{pout})-(\ref{pin}), in terms of the neutrino energy, $E_\nu$, using energy conservation, i.e., $Q_{ij} = E_{e} + E_{\nu}$ for beta decay, and $Q_{ij} + E_{e} = E_{\nu}$ for electron capture.  These phase space integrals contain all of the dependence on neutrino energy, and thus solely determine the shape of the \n\ energy spectra.

Following the approach of \citet{Langanke:2001b}, we adopt a single, effective $Q$-value, $Q$,  for each isotope, and treat it as a fit parameter.  This effective \qv\ is found by requiring that the average \n\ energy matches the value obtained from the nuclear rate tables, i.e.: 
\begin{equation}
\langle E_{\nu,\overline{\nu}}\rangle = \frac{\int_{0}^{\infty }{\left( d\lambda/dE_{\nu} \right) E_{\nu} dE_{\nu}}}{\int_{0}^{\infty}{\left( d\lambda/dE_{\nu} \right) dE_{\nu}}} = \frac{{\mathcal E}^{\nu,\overline{\nu}}}{\lambda^{EC,PC} + \lambda^{\beta^{\pm}}}~,
\label{eq:aveE}
\end{equation}
where the quantities on the right side of the equation are obtained from the rate tables.  Here $\lambda^{EC,PC}$ and $\lambda^{\beta^{\pm}}$ are the electron (positron) capture rate and $\beta^{\pm}$ decay rates \citep{Langanke:2001}; $\lambda = \lambda^{EC,PC} + \lambda^{\beta^{\pm}}$, and ${\mathcal E}^{\nu,\overline{\nu}}$ is the rate of energy loss as (anti-) neutrinos.  Note that $\langle E_{\nu,\overline{\nu}}\rangle$ is a combined value, including both the capture and decay values weighted by the respective rates.  Therefore, by construction, the $Q$-value found from Eq. (\ref{eq:aveE}) is the same for both decay and capture. Here and throughout the paper, subscripts or superscripts such as in $\langle E_{\nu,\overline{\nu}}\rangle$ indicate that an equation is true for, in this example, $\langle E_{\nu}\rangle$ and $\langle E_{\overline{\nu}}\rangle$ with all subscripted values in the equation taking either $\nu$ or $\overline{\nu}$ as necessary.  

Combining Eqs. (\ref{eq:weakDecayRate})-(\ref{eq:aveE}), we find the spectra for the weak processes for a single isotope: 
\begin{eqnarray}
\phi_{EC,PC} &=& N_{EC,PC} \frac{E_{\nu}^{2}(E_{\nu} - Q)^{2}}{1 + \exp{((E_{\nu} - Q - \mu_{e})/kT)}}  \Theta(E_{\nu} - Q - m_{e}) \label{eq:phiEC}\\
\phi_{\beta} &=& N_{\beta} \frac{E_{\nu}^{2}(Q - E_{\nu})^{2}}{1 + \exp{((E_{\nu} - Q + \mu_{e})/kT)}} \Theta(Q - m_{e} - E_{\nu})~, \label{eq:phiBeta}
\end{eqnarray}
where $N_i$ is a normalization factor defined such that
\begin{eqnarray}
\lambda^{i} = \int_{0}^{\infty} \phi_{i} dE_{\nu} \,\,\,\,\,\,\, i = EC, PC, \beta^{\pm}. \label{eq:phiNorm}
\end{eqnarray}
Let us note that the \ps\ environment is different from that of a \sn: for a \sn\ the high electron degeneracy inhibits beta decay, so that electron capture plays a stronger role  \citep{Langanke:2001b, Sullivan:2015}; in our case of interest, instead, lower degeneracies allow $\beta$  decays to proceed.  Their importance has been emphasized in \citep{Heger:2001b, MartinezPinedo:1998, Aufderheide:1994b, Aufderheide:1994}. 

Finally, the total $\nue$ (or $\barnue$) spectrum is found by a weighted sum over all the isotopes present:
\begin{equation}
\Phi_{\nu,\overline{\nu}} =\sum_{k}  \phi_k n_k = \sum_{k} X_{k} \phi_{k} \frac{\rho}{m_{p}A_{k}}~,
\label{eq:PhiTotal}
\end{equation}
where $\rho$ is the mass density and $m_{p}$ is the mass of the proton. Here $\phi_{k}$ is the sum of the normalized electron (positron) capture
 and $\beta^{\pm}$ decay spectra for isotope $k$; $n_k=X_{k}\rho/(m_{p}A_{k})$ is the number density of the same isotope, and $X_{k}$, $A_k$  are its mass fraction and atomic number respectively \citep{Odrzywolek:2009yl}. 
 
 The values of $X_{k}$ and $\rho$ are taken from MESA calculations \citep{Paxton:2010ig, Paxton:2013jy, Paxton:2015vl}.  
The features of  the spectrum $\Phi_{\nu,\overline{\nu}}$ depend on the temperature, density, and isotopic abundances.  For the center of a star  immediately before collapse (T $\approx 4-5 \times10^{9}$ K, $\rho \approx 10^{7}$ g/cm$^{3}$), the total spectrum can extend to several MeV.

\subsection{Thermal Processes}
\label{sec:thermal}

Let us now discuss the three most important thermal processes: plasmon decay, photoneutrino production, and pair annihilation (Table \ref{tab:microphys}).  The total emissivities of all these processes, over a range of temperatures and densities, were discussed in detail by \citet{Itoh:1983, Itoh:1989, Itoh:1992ss, Itoh:1996im, Itoh:1996vq} . The differential rates and emissivities of selected process have been discussed by several authors \citep{Odrzywolek:2007kv, Dutta:2004ug, Ratkovic:2003pu, Misiaszek:2006jo, Kato:2015ec, Asakura:2015sw}.  In this section, we summarize the formalism relevant to our calculation.

\subsubsection{Plasma Neutrino Process}\label{sec:plasma}

In the plasma neutrino process, an excitation in the plasma (plasmon) decays into a neutrino-antineutrino pair.  As shown in \citet{Itoh:1996vq}, plasma neutrinos dominate the total emissivity at high densities.  Detailed derivations and discussions of this process are given in \citet{Ratkovic:2003pu,Odrzywolek:2007kv}. Drawing from this literature, here the essential equations for calculating the plasmon decay neutrino spectrum are summarized.  

The total rate $R$ and emissivity $Q$ are given by the integrals \citep{Ratkovic:2003pu}
\begin{eqnarray}
R & =& \sum_{\epsilon} \int \frac{d^{3}k}{2\omega (2\pi)^{3}} Z(k) \frac{d^{3}q_{1}}{2 \mathcal{E}_{1} (2\pi)^{3}} \frac{d^{3}q_{2}}{2 \mathcal{E}_{2} (2\pi)^{3}} \left[ \langle | \mathcal{M} |^{2} \rangle f_{\gamma^{\star}}(\omega) (2\pi)^{4}\delta^{4}(K - Q_{1} - Q_{2}) \right] \label{eq:totRatePlasma} \\ 
Q &=&  \sum_{\epsilon} \int \frac{d^{3}k}{2\omega (2\pi)^{3}} Z(k) \frac{d^{3}q_{1}}{2 \mathcal{E}_{1} (2\pi)^{3}} \frac{d^{3}q_{2}}{2 \mathcal{E}_{2} (2\pi)^{3}}\left[ (\mathcal{E}_{1} + \mathcal{E}_{2}) \langle | \mathcal{M} |^{2} \rangle f_{\gamma^{\star}}(\omega) (2\pi)^{4}\delta^{4}(K - Q_{1} - Q_{2}) \right]
\label{eq:totEmissPlasma} ~,
\end{eqnarray}
where $Q_{1,2} = (\mathcal{E}_{1,2}, \bf{q_{1,2}})$ are the four-momenta of the daughter \n\ pair;  $K = (\omega, \bf{k})$ and $f_{\gamma^{*}}$ are the plasmon four-momentum and spectrum:
\begin{equation}
f_{\gamma^{*}} = \frac{1}{e^{\omega_{T,L}/kT} - 1}. 
\label{eq:fgamma}
\end{equation}
The factor $Z(k)$ in Eqs. (\ref{eq:totRatePlasma}) and (\ref{eq:totEmissPlasma}) is the residue from integrating around the pole in the propagator.  The sums in Eqs. (\ref{eq:totRatePlasma}) and (\ref{eq:totEmissPlasma}) are over the polarizations appropriate for the decay mode.  There are two possible decay modes: transverse (T), which has two polarizations, and longitudinal (L), with one polarization.  

The term $\langle | \mathcal{M} |^{2} \rangle$ is the squared matrix element for the process.  The effective vertex for plasmon decay has both vector and axial vector pieces \citep{Braaten:1993hf}.  For the longitudinal decay mode, the axial vector term disappears, leaving the squared matrix element as \citep{Ratkovic:2003pu}: 
\begin{eqnarray}
\langle | \mathcal{M} |^{2} \rangle^{L} & = & 2 \frac{G_{F}^{2} (C_{V}^{f})^{2}}{\pi \alpha} \left( \frac{\omega_{L}^{2} - k^{2}}{k^{2}}\right)^{2} \Pi_{L}^{2}(\omega_{L}, k) \times \left[ \frac{(\mathcal{E}_{1}\omega_{L} - {\bf q}_{1}\cdot {\bf k})(\mathcal{E}_{2}\omega_{L} - {\bf q}_{2}\cdot {\bf k})}{\omega_{L}^{2} - k^{2}} + \frac{({\bf k} \cdot{\bf q}_{1})({\bf k} \cdot{\bf q}_{2})}{k^{2}} - \frac{\mathcal{E}_{1}\mathcal{E}_{2} + {\bf q}_{1}\cdot{\bf q}_{2}}{2}\right].\label{eq:MLPlasma}
\end{eqnarray}
On the other hand, both vector and axial vector pieces survive in the calculation of the transverse decay mode.  After squaring, we are left with a transverse vector term proportional to $(C_{V}^{f})^{2}$, an axial term with coefficient $(C_{A}^{f})^{2}$, and a mixed term with a factor $(C_{A}^{f}C_{V}^{f})$ \citep{Ratkovic:2003pu}.  Put together, the squared matrix element for the transverse decay mode is 
\begin{eqnarray}
\langle | \mathcal{M} |^{2} \rangle^{T} & = &\frac{G_{F}^{2}}{\pi \alpha}\left[ \left((C_{V}^{f})^{2} \Pi_{T}^{2}(\omega_{T}, k) +  (C_{A}^{f})^{2}\Pi_{A}^{2}(\omega_{T}, k)\right)\right. \nonumber \\
& & \left. \times \left( \mathcal{E}_{1}\mathcal{E}_{2}  - \frac{({\bf k} \cdot{\bf q}_{1})({\bf k} \cdot{\bf q}_{2})}{2}\right) + 2 (C_{A}^{f}C_{V}^{f}) \frac{\Pi_{A}(\omega_{T},k)\Pi_{T}(\omega_{T},k)}{k}  \times \left( \mathcal{E}_{1}({\bf k} \cdot{\bf q}_{2}) - \mathcal{E}_{2}({\bf k} \cdot{\bf q}_{1})  \right) \right]. \label{eq:MTPlasma}
\end{eqnarray}
The functions $\Pi_{L,T,A}$ are the longitudinal, transverse, and axial polarization functions, which are defined in \citep{Ratkovic:2003pu}.  The total emissivity for the plasmon decay process is found by summing all of these channels. The vector and axial couplings, $C_{V}^{f}$ and $C_{A}^{f}$, are 
\begin{eqnarray}
C_{V}^{e} & = & \frac{1}{2} + 2 \sin^{2}(\theta_{W}) \nonumber \\
 C_{A}^{e} & = & \frac{1}{2} \nonumber \\
C_{V}^{x} & = & C_{V}^{e} - 1 \nonumber \\
C_{A}^{x} & = & C_{A} - 1,
\end{eqnarray}
with $\sin^{2}(\theta_{W})$ = 0.226.   The difference in these couplings results in a suppression of the $\nux$ flavors by factors of $(C_{V}^{x}/C_{V}^{e})^{2} \approx 3\times10^{-3}$ \citep{Odrzywolek:2007kv}.  

After integrating Eqs. (\ref{eq:totRatePlasma}) and (\ref{eq:totEmissPlasma}) over the plasmon momentum and  the angle between the outgoing neutrinos, one gets the rate, differential in the \n\ energy $\mathcal{E}_{1}$  \citep{Odrzywolek:2007kv}: 
\begin{equation}
\frac{dR^{L,T}}{d\mathcal{E}_{1}} = \int_{0}^{\infty}{d\mathcal{E}_{2}\frac{g_{L,T}}{\pi^{4}} Z_{L,T} \langle | \mathcal{M} |^{2} \rangle^{L,T} f_{\gamma^{*}} J_{L,T} S}~, 
\label{diffrateplasmon}
\end{equation}
with $g_{T} = 2$ and $g_{L} = 1$ accounting for the number of polarizations for each mode.  The new factor $J_{L,T}$ is the Jacobean resulting from the $\delta$ function integration.  The factor $S$, is a product of step functions, describing the physically relevant region: 
\begin{equation}
S = \Theta(4\mathcal{E}_{1}\mathcal{E}_{2} - m_{L,T}^{2}) \Theta(\mathcal{E}_{1} + \mathcal{E}_{2} - \omega_{L,T}) \Theta(\omega_{max} - \mathcal{E}_{1} - \mathcal{E}_{2}).
\end{equation}

The maximum plasmon energy, $\omega_{max}$, is finite for longitudinal plasmons, and depends on the temperature and density of the environment.  Instead, $\omega_{max} \rightarrow \infty$ for transverse plasmons.  We have used the Braaten-Segel approximations \citep{Braaten:1993hf} in all of these calculations, allowing the differential rate to be calculated analytically.  The expressions for various plasma parameters, such as $\omega_{max}$, $Z_{L,T}$, $\Pi_{L,T,A}$, and $J_{L,T}$ in this approximation are given in \cite{Ratkovic:2003pu,Odrzywolek:2007kv}.  The total emissivity calculated through this method is consistent with the Itoh \textit{et al.} formula for the plasma process.  

The spectra of \ns\ from plasmon decay are typically colder that those from other processes (see figs. \ref{spectra15} - \ref{spectra30}).  The neutrinos resulting from longitudinal plasmon decay are limited by $\omega_{max}$, with $\omega_{max} < \mathrm{MeV}$ for  typical \ps\ temperatures and densities.  Neutrinos from transverse plasmon decay have no such energy restriction, however, and can extend beyond 1 MeV in some cases. Probably, the plasmon decay contribution can not be individually identified in a detector; nevertheless it can have a  major impact on the total neutrino emissivity at some points in the lifetime of the star.

\subsubsection{Photoneutrino Process}
\label{sec:photo}

For the photoneutrino process, we follow the extensive discussion in \citet{Dutta:2004ug}.  The calculation of rates and emissivities is very similar to that for the plasma neutrino process.  In this case, the total rate $R$ and emissivity $Q$ are calculated by performing the integrals
\begin{eqnarray}
R & =& \int \frac{2 d^{3}p}{(2\pi)^{3}}\frac{f_{e}(E_{P})}{2E_{P}}\int \frac{\xi d^{3}k}{(2\pi)^{3}} \frac{f_{\gamma}(\omega)}{2\omega}  \nonumber \\
& & \times \int \frac{d^{3}p'}{(2\pi)^{3}} \frac{1-f_{e}(E_{P'})}{2E_{P'}}\int\frac{d^{3}q_{1}}{(2\pi)^{3}}\frac{1}{2\mathcal{E}_{1}} \nonumber \\
& & \times \int\frac{d^{3}q_{2}}{(2\pi)^{3}}\frac{1}{2\mathcal{E}_{2}} (2\pi)^{4}\delta^{4}(P + K - P' - Q_{1} - Q_{2})\nonumber \\
& &\times \frac{1}{\zeta} \langle | \mathcal{M} |^{2} \rangle\label{eq:totPhotoRate}\\ 
Q &=& \int \frac{2 d^{3}p}{(2\pi)^{3}}\frac{f_{e}(E_{P})}{2E_{P}}\int \frac{\xi d^{3}k}{(2\pi)^{3}} \frac{f_{\gamma}(\omega)}{2\omega} \nonumber\\
& & \times \int \frac{d^{3}p'}{(2\pi)^{3}} \frac{1-f_{e}(E_{P'})}{2E_{P'}}\int\frac{d^{3}q_{1}}{(2\pi)^{3}}\frac{1}{2\mathcal{E}_{1}}\nonumber\\
& & \times \int\frac{d^{3}q_{2}}{(2\pi)^{3}}\frac{1}{2\mathcal{E}_{2}} (2\pi)^{4}\delta^{4}(P + K - P' - Q_{1} - Q_{2})\nonumber\\
& &\times(\mathcal{E}_{1} + \mathcal{E}_{2}) \frac{1}{\zeta} \langle | \mathcal{M} |^{2} \rangle\label{eq:totPhotoEmiss}~. 
\end{eqnarray}
As in Eqs. (\ref{eq:totRatePlasma}) and (\ref{eq:totEmissPlasma}), these expressions are integrals of the squared matrix element over the incoming and outgoing momenta, taking into account the photon and electron distributions and energy conservation.  Here, $P = (E_{P},{\bf{p}_{1,2}})$ is the four momentum for the incoming electron and $P' = (E_{P'},{\bf{p'}})$ is the same for the outgoing electron.  Following the notation defined in the plasma neutrino discussion, $Q_{1,2} = (\mathcal{E}_{1,2}, {\bf q_{1,2}})$ are the four momenta for the outgoing neutrino pair and $K = (\omega, {\bf k})$ is the four momentum of the photon.  

The sums run over the polarization of the photon and spin of the incoming and outgoing electrons. Here the initial factor of 2 accounts for the spin of the incoming electron, $\xi$ is due to the polarization of the incoming photon, and $\zeta$ from the spins of the outgoing electron and neutrinos.  Similar to plasmon decay, there are transverse and longitudinal modes for the photoneutrino process.  For the transverse mode of the photon, $\xi = 2$ and $\zeta = 4$.  In the longitudinal case, $\xi = 1$ and $\zeta = 2$.  \citep{Dutta:2004ug}.  

As discussed in \citet{Dutta:2004ug}, the squared matrix element can be derived to be 
\begin{eqnarray}
\langle | \mathcal{M} |^{2} \rangle^{T(L)}& = & 32 e^{2} G_{F}^{2} \left[ \left( (C_{V}^{f})^{2} - (C_{A}^{f})^{2} \right) m_{e}^{2} \mathcal{M}_{-}^{T(L)}  + \left( (C_{V}^{f})^{2} + (C_{A}^{f})^{2} \right) \mathcal{M}_{+}^{T(L)}  + C_{V}^{f} C_{A}^{f} \mathcal{M}_{\times}^{T(L)} \right] \nonumber\\
 & & \label{eq:photoM2}
\end{eqnarray}
This simple form includes the terms $\mathcal{M}_{-}$, $\mathcal{M}_{+}$, and $\mathcal{M}_{\times}$.  These terms are complicated combinations of products of the four-momenta and photon polarizations.  The full expressions can be found in the Appendix A of \cite{Dutta:2004ug}.  

The delta function in Eqs. (\ref{eq:totPhotoRate}) and (\ref{eq:totPhotoEmiss}) can be used to complete the integration over the incoming electron momentum ${\bf p}$ and the incoming photon angle $\theta_{k}$. As in \citet{Dutta:2004ug}, here the coordinate system is such that one neutrino momentum is aligned with the z-axis, with the second neutrino momentum in the x-z plane at an angle $\theta_{q_{1}q_{2}}$, while the outgoing electron momentum ${\bf p'}$ is in an arbitrary direction defined by angles $\theta_{e}$ and $\phi_{e}$.  In this formalism, we can find the total four-momentum of the system, $P_{tot} = P + K = P' + Q_{1} + Q_{2}$, and thus determine the momenta of the incoming photon and electron.  For details on these initial integrations, we refer you to \citet{Dutta:2004ug}.

The result for the differential rates and emissivities is the following: 
\begin{eqnarray}
\frac{dR}{d\mathcal{E}_{1}} &=& \frac{\pi^{2}}{(2\pi)^{9}} \mathcal{E}_{1} \int_{0}^{\infty}\mathcal{E}_{2} d\mathcal{E}_{2} \int_{-1}^{1}d\cos{\theta_{q_{1}q_{2}}} \nonumber \\
& & \times \int_{0}^{\infty} \frac{|{\bf p'}|^{2}}{E_{P'}} d|{\bf p'}| \int_{-1}^{1}d\cos{\theta_{e}} \nonumber \\
& & \times \int_{0}^{2\pi}d\phi_{e} \left[ 1 - f_{e}(E_{P'}) \right] I(p', q_{1}, q_{2}) \label{eq:diffPhotoRate}\\ 
\frac{dQ}{d\mathcal{E}_{1}} &=& \frac{\pi^{2}}{(2\pi)^{9}} \mathcal{E}_{1} \int_{0}^{\infty}\mathcal{E}_{2} d\mathcal{E}_{2} \int_{-1}^{1}d\cos{\theta_{q_{1}q_{2}}} \nonumber \\
& & \times \int_{0}^{\infty} \frac{|{\bf p'}|^{2}}{E_{p'}} d|{\bf p'}| \int_{-1}^{1}d\cos{\theta_{e}} \nonumber \\
& & \times \int_{0}^{2\pi}d\phi_{e} \left[ 1 - f_{e}(E_{P'}) \right] \nonumber\\
& &  (\mathcal{E}_{1} + \mathcal{E}_{2}) I(p', q_{1}, q_{2}) \label{eq:diffPhotoEmiss}. 
\end{eqnarray}
where the integral $I(p', q_{1}, q_{2})$ is defined as
\begin{eqnarray}
I(p',q_{1},q_{2}) & = & \frac{1}{2\pi^{2}} \int_{0}^{\infty} \frac{|{\bf k}|}{\omega} d|{\bf k}| \int_{0}^{2\pi} d\phi_{k} \left[ f_{\gamma}(\omega) f_{e}(E_{P}) \frac{1}{|\bf{P}_{tot}|} \langle | \mathcal{M} |^{2} \rangle^{T(L)}\right]. \label{eq:photoIntI} 
\end{eqnarray}
%
The seven-dimensional integrals in Eqs. (\ref{eq:diffPhotoRate})-(\ref{eq:diffPhotoEmiss}) are calculated by a Monte Carlo method, as in \citet{Dutta:2004ug}.  As with the plasma process, our calculation using the methods of \citet{Dutta:2004ug} are consistent with the formula from \citet{Itoh:1996im}.

As will be shown in Section \ref{sec:results}, photoneutrinos dominate the total emissivity at a few locations in the outer shells of the star, in agreement with the results in \citet{Itoh:1996im}.  However, for the part of the \ps\ \n\ flux that is of most interest for detection -- corresponding to higher temperatures and densities --  these neutrinos are typically overwhelmed by pair and $\beta$ process neutrinos.

\subsubsection{Pair Annihilation Neutrinos}

The emissivity for \ns\ from pair annihilation, $e^{+} + e^{-} \rightarrow \nu + \overline{\nu}$, can be calculated similarly to Eqs. (\ref{eq:totRatePlasma}) and (\ref{eq:totEmissPlasma}): the squared matrix element for this process has to be integrated over the incoming and outgoing momenta, including considerations for the electron and positron momentum distributions and energy conservation.  For the pair annihilation process, the expressions for the total rate $R$ and emissivity $Q$ are
\begin{eqnarray}
R & = & \frac{4}{(2\pi)^{8}} \int{ \frac{d^{3}p_{1}}{2E_{1}} f_{1}(E_{1})}  \int{\frac{d^{3}p_{2}}{2E_{2}} f_{2}(E_{2})} \int{ \frac{d^{3}q_{1}}{2\mathcal{E}_{1}} } \int{\frac{d^{3}q_{2}}{2\mathcal{E}_{2}} } \left[ 2  \langle | \mathcal{M} |^{2} \rangle \delta^{4}\left( P_{1} + P_{2} - Q_{1} - Q_{2}\right) \right], \\
Q & = &  \frac{4}{(2\pi)^{8}} \int{ \frac{d^{3}p_{1}}{2E_{1}} f_{1}(E_{1})}  \int{\frac{d^{3}p_{2}}{2E_{2}} f_{2}(E_{2})} \int{ \frac{d^{3}q_{1}}{2\mathcal{E}_{1}} } \int{\frac{d^{3}q_{2}}{2\mathcal{E}_{2}} } \left[ \left(\mathcal{E}_{1} + \mathcal{E}_{2}\right)  \langle | \mathcal{M} |^{2} \rangle \delta^{4}\left( P_{1} + P_{2} - Q_{1} - Q_{2}\right) \right]. 
\end{eqnarray}
where the squared matrix element, appropriately averaged over spin, is
\begin{eqnarray}
\langle | \mathcal{M} |^{2} \rangle = 8 G_{F}^{2} \left( \left( C_{A}^{f} - C_{V}^{f}\right)^{2}\left(P_{1}\cdot Q_{1}\right) \left( P_{2}\cdot Q_{2}\right) \right. \nonumber \\
+ \left( C_{A}^{f} + C_{V}^{f}\right)^{2}\left(P_{2}\cdot Q_{1} \right) \left(P_{1}\cdot Q_{2} \right)
 \nonumber  \\ \left.+ m_{e}^{2} \left( C_{V}^{2} - C_{A}^{2}\right) \left(Q_{1}\cdot Q_{1}\right)\right),
\end{eqnarray}
where $P_{1,2} = (E_{1,2}, {\bf p}_{1,2})$ are the four-momenta of the incoming electron and positron, and $Q_{1,2} = (\mathcal{E}_{1,2}, {\bf q}_{1,2})$ are the four-momenta of the outgoing neutrino and antineutrino. The functions $f_{i}(E_{i})$ are the Fermi-Dirac distributions for the electron and positron.

As in the previous derivations for plasmon decay and the photoneutrino process, the delta function can be used to simplify the integral.  Extensive algebra, an example of which can be found in \citet{Hannestad:1994}, reduces the calculation of the differential rate to a three dimensional integral over the magnitudes of the electron and positron momenta and the angle between them:
\begin{equation}
\frac{dR}{d\mathcal{E}_{i}} = \int d^{3}{p}_{1}d^{3}{p}_{2}\frac{d\sigma \varv}{d\mathcal{E}_{i}} f_{1}f_{2},
\end{equation}
where 
\begin{equation}
d\sigma \varv = \frac{1}{2E_{1}} \frac{1}{2E_{2}} \frac{1}{(2\pi)^{2}} \delta^{4}(P_{1} + P_{2} - Q_{1} - Q_{2}) \frac{d^{3}q_{1}}{2 \mathcal{E}_{1}} \frac{d^{3}q_{2}}{2 \mathcal{E}_{2}}   \langle | \mathcal{M} |^{2} \rangle~ ,
\end{equation}
and $\varv$ being the relative velocity of the electron-positron pair.  We have performed this integral using Monte Carlo integration.  

In agreement with the results of \citet{Itoh:1996im}, the results of MESA show (Section \ref{sec:results}) that pair neutrinos are the major contributors to the neutrino flux from the center of the star where temperatures and densities are highest.  The energies of these neutrinos can reach up to several MeV and, along with the $\beta$ process, produce most of the potentially detectable neutrinos.  

\section{Results: neutrino emission in an evolving star}
\label{sec:results}

 \subsection{The calculation: technical aspects, inputs and outputs}
\label{code}

We employed the stellar evolution code MESA, version r7624 \citep{Paxton:2010ig, Paxton:2013jy, Paxton:2015vl}, to simulate the evolution of progenitor stars with masses $M=15, 20, 25, 30~ M_\odot$, from pre-main-sequence (pre-MS) to the onset of core collapse, which is defined as when the infall velocity anywhere in the star exceeds  $V_{max}=1\times10^8 {\rm cm~s^{-1}}$.   This final instant is defined as $t=t_c=0$, and all the earlier times $t$ ($t<0$) will be defined relative to it, so that $-t >0$ will indicate the time-to-collapse. The progenitor models used here are single, non-rotating, non-mass losing stars with a solar metallicity (i.e., mass fraction $Z=0.02$ of elements heavier than He)  and a solar abundance distribution from \citet{grevesse_1998_aa}; see \citet{farmer_2016_aa} for more details. Note that the range of masses we consider covers some of the diversity expected in the final outcome of the collapse: while the progenitors with lower mass are likely to generate a strong shockwave, resulting in a robust supernova explosion, the  heavier ones ($M=25, 30~ M_\odot$) were found to be candidates for direct black hole formation (without explosion, a ``failed supernova"), due to their greater compactness (see, e.g. \cite{OConnor:2010moj,Pejcha:2014wda}).

The MESA simulation includes the effects of semiconvection, convective overshooting and thermohaline   mixing. We use the models from \citet{farmer_2016_aa} with
$\maxdm=0.1$, where \maxdm\ specifies the maximum cell mass, and MESA's $\mesh=1.0$, where \mesh\ controls the relative variance between cells.  The combination of these two settings results in $\approx 1000 - 2000$ spatial zones at core collapse.   The isotopic composition of the star was modeled using MESA's \net\  reaction network which includes 204 isotopes up to $^{66}$Zn, including all relevant reactions, fully coupled to the hydrodynamics from the pre-MS to core collapse. The MESA inlists and stellar models are publicly available \footnote{\url{http://mesastar.org/results}}. 

For each time step of the evolution, MESA produces in output the temperature, mass density, proton fraction, electron degeneracy, and isotopic composition as a function of the radial coordinate. Using formulae from \citet{Itoh:1983, Itoh:1989, Itoh:1992ss, Itoh:1996im, Itoh:1996vq}, MESA also calculates, and provides in output, the \n\ energy emissivity $Q$ (i.e., the total energy emitted in \ns\ per unit volume per unit time) for each production channel in Table \ref{tab:microphys}.  This quantity was useful as a consistency check for the semi-analytical formalism in Sec. \ref{sec:production}, which reproduces it correctly. The same formalism was then used to perform a separate calculation (which uses the thermodynamical quantities calculated in MESA, but is not embedded in it) of the \n\ energy spectra for each of the production channels. 

To emphasize the part of the \n\ flux that is potentially detectable,  from the \n\ spectra we derived a partial energy emissivity, $\qth$, defined as the energy emitted in neutrinos with energy $E > E_{th}$ per unit volume per unit time, with $E_{th}=2$ MeV being an indicative threshold for detectability.  This is a realistic  value for liquid scintillator detectors (e.g., \citep{An:2015jdp}); the threshold is typically higher than $\sim 5$ MeV at liquid argon and water Cherenkov experiments \citep{cdr,Abe:2011ts}. 

The following subsections illustrate our results in graphics and text.  The reader is referred to Table \ref{pointstab} for the complete numerical details. 

\subsection{A neutrino history: emissivity profiles}
\label{history}

\begin{figure*}
\includegraphics[width =\linewidth]{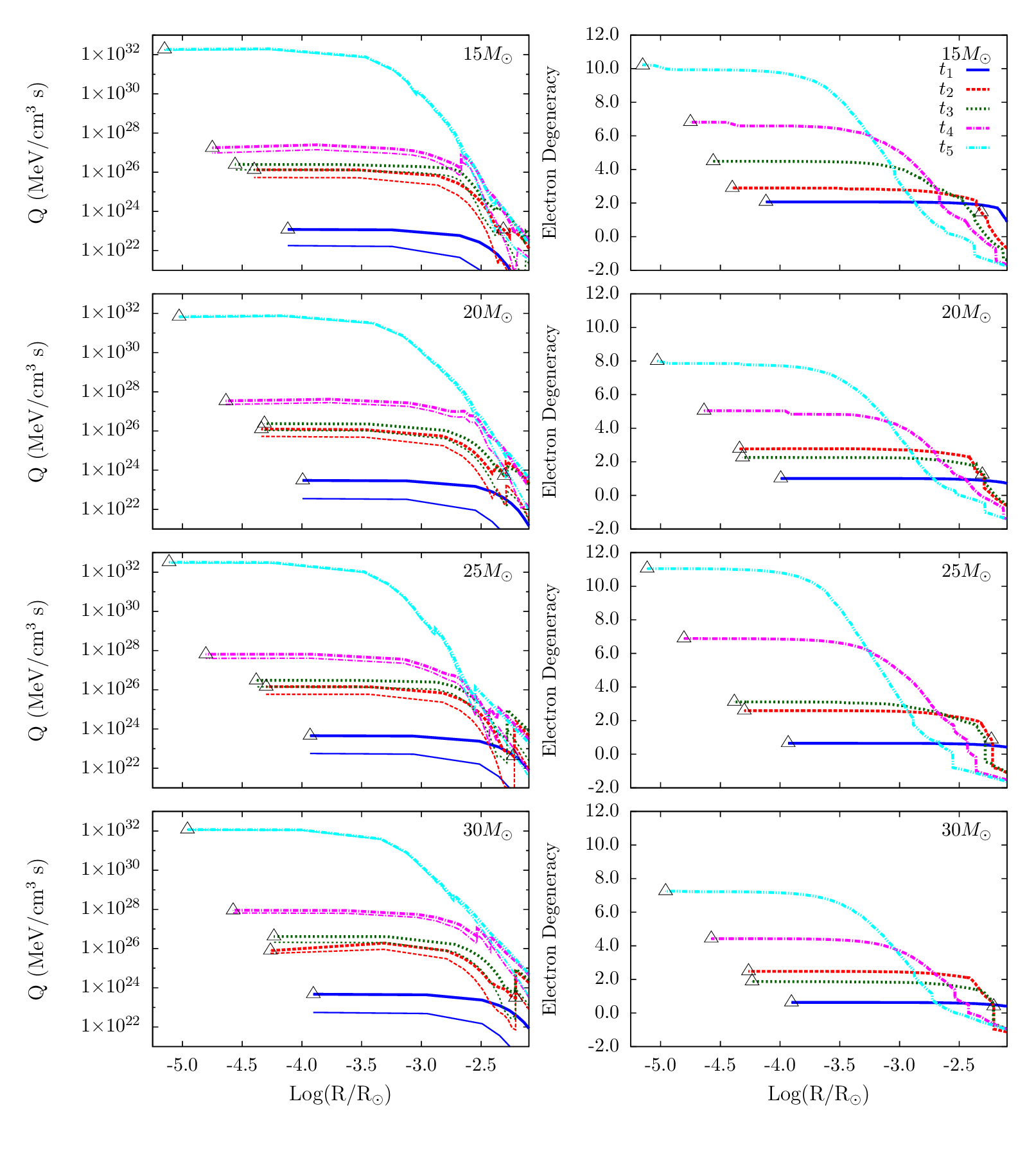}
\caption{
{\it Left}:  The total  \n\ emissivity in the star, $Q$ (thick lines), and the emissivity for $E>2$ MeV, $Q_{th}$ (thin lines), as functions of the radial coordinate, for each progenitor star, and for the selected times in Table \ref{pointstab}.  {\it Right}:  The radial profiles of the electron degeneracy parameter, $\eta_e=\mu_e/T$, for the same progenitors and times. 
In all figures, sample points are marked; details about them are given in Table \ref{pointstab} and Figs. \ref{spectra15} - \ref{spectra30}. }
\label{evolution} 
\end{figure*}
Let us first trace the time evolution of a star in the  \ps\ phase in terms of \n\ emission.  We focus on the electron neutrino species, and do not include \n\ flavor oscillations. The $\mu$ and $\tau$ flavors (collectively denoted $\nux$ here) only give subdominant contributions to rates and emissivities, because they are not produced in $\beta$ processes, and are suppressed in thermal processes.  The $\mu$ and $\tau$ flavors are included in the total emissivities, but their spectra are not shown in figs. \ref{spectra15} - \ref{spectra30}.

It is known (see e.g., \cite{1978ApJ...225.1021W}) that the rate of evolution, and in particular the duration of each stage of nuclear burning, depends strongly on the progenitor mass, with more massive stars evolving more rapidly. For this reason, to facilitate comparisons between runs with different progenitors, for each star model results have been generated at five selected times, $t_n~(n=1,....,5)$ (see Table \ref{pointstab}),  which were chosen to correspond to a physical event or phase in the evolution: at $t_1$, the star's core is at the beginning of the O-burning phase; $t_{2}$ is  approximately central though the core Si-burning phase, and $t_{3}$ marks the end of it. We set $t_5$ to be the last step of the evolution, corresponding to the onset of collapse: $t_5=t_c$.  At $t_{5}$, Si burning is occurring in an outer shell, around $\log{(R/R_{\odot})} \approx -2.5$.  

As expected, the $t_n$ are very different for the different progenitor models: 
for $M=15~M_\odot$ we find $t_1 \simeq - 1.4 \times 10^4$ hr (i.e., about 600 days before collapse), while for  $M=30~M_\odot$ we have $t_1 \simeq - 1.0 \times 10^3$ hr ($\sim$44 days).  Likewise, we find $t_3\simeq -7.4$ hr for the $M=15~M_\odot$ and  $t_3= - 0.97$ hr, for $M=30~M_\odot$ star, thus confirming the faster evolution of more massive progenitors. 

As an exception, the time $t_4$ ($t_3 < t_4 <t_5$) has been specifically set to be the same ($t_4 \sim -0.5$ hrs) for all stars, so to offer guidance on how strongly an observed \ps\ \n\ time profile might depend on the progenitor mass. 

 Fig. \ref{evolution} shows the radial distribution of $Q$ and $\qth$, at each of the selected times and for each progenitors model.   At all times, $Q$ is maximum in the region $R \lta 10^{-2.5}~R_\odot$, and declines roughly as $R^{-6}$ for larger radii.   $\qth$ is within an order of magnitude or so of $Q$ in the central region, and falls more steeply than $Q$ with increasing radius.  We note sharp, time-dependent discontinuities in the emissivities, which reflect the shell structure of the star.  Fig. \ref{evolution} also shows the radial profiles of the electron degeneracy parameter, $\eta_e = \mu_e/T$.  It appears clearly that $\eta_e$ increases strongly over time in the star's core, rising from $\eta_e \sim 0$ to values as high as $\eta_e \simeq 11$ at $t=t_5$.   As a consequence, $\nue$ production through $\beta^-$ decays becomes increasingly inefficient due to electron Pauli blocking (see Sec.~\ref{spectraresults}).   

When comparing results for different stellar progenitors (Fig. \ref{evolution}, Table \ref{pointstab}), it appears that differences in the emissivities are larger at early times, and become more modest at later times, with the more massive stars generally having larger \n\ emissivities.  For example, the emissivity integrated over the volume of the star, $Q_{int}$, at  time $t_1$ is a factor of $\sim 20$ larger for the $M=30 \msun$ star than for the $15 \msun$ one, but at $t=t_5$, differences in $Q_{int}$ are at the level of tens of per cent only.  The local emissivities $Q$ at selected radii inside the star show a similar trend (Table \ref{pointstab}), however differences between progenitors are more modest, at the level of a factor of $\sim 2-3$.  
At the fixed time $t_4$ (half hour before collapse), differences are substantial, about an order of magnitude in $Q_{int}$ and a factor $\sim 4$ in $Q$ in the core, between the least and the most massive progenitors, thus suggesting the possibility to use the time profile of a \n\ signal in a detector for progenitor identification.  
The dependence on the progenitor mass/model found here is overall consistent with the faster evolution of the more massive progenitors.  The electron degeneracy, $\eta_e$, only varies mildly between progenitors, by tens of per cent. 

 We note that variations are not always monotonic with the progenitor mass; for example the $M=20 \msun$ star has the largest emissivity and the largest $\eta_e$ in the core (Fig. \ref{evolution}).  This non-monotonic behavior reflects the underlying non-monotonic character of the stellar models themselves and the non-linear nature of stellar evolution.  As a star evolves off of the the giant branch, forming a carbon/oxygen core, the degeneracy of the core begins to play a larger role in the stars evolution. The relevant timescales are set by an interplay of the neutrino losses and the ignition of each subsequent fuel source. A star may ignite one or more off center shell flashes while burning oxygen or silicon in its core. If the energy released by the ignition is much greater than the local neutrino cooling rate, then the shell flash may drive a convection zone, which can mix in fresh fuel keeping the shell burning.  This can lead to discontinuous changes (with respect to changes in the initial mass) between models that do and do not drive a convection zone \citep{Heger:2001b}. This complex interplay between the type of ignitions, formation and size of convection zones, and the composition of the material post ignition, can lead to non-linear behavior of the star with respect to increasing
initial mass. For instance at $\approx 20\msun$ there is a complex transition between convective carbon burning
and radiative carbon burning \citep{timmes_1996_ac,heger_2000_aa,Hirschi_2004_aa}.  

\begin{figure*}
\includegraphics[width =\linewidth]{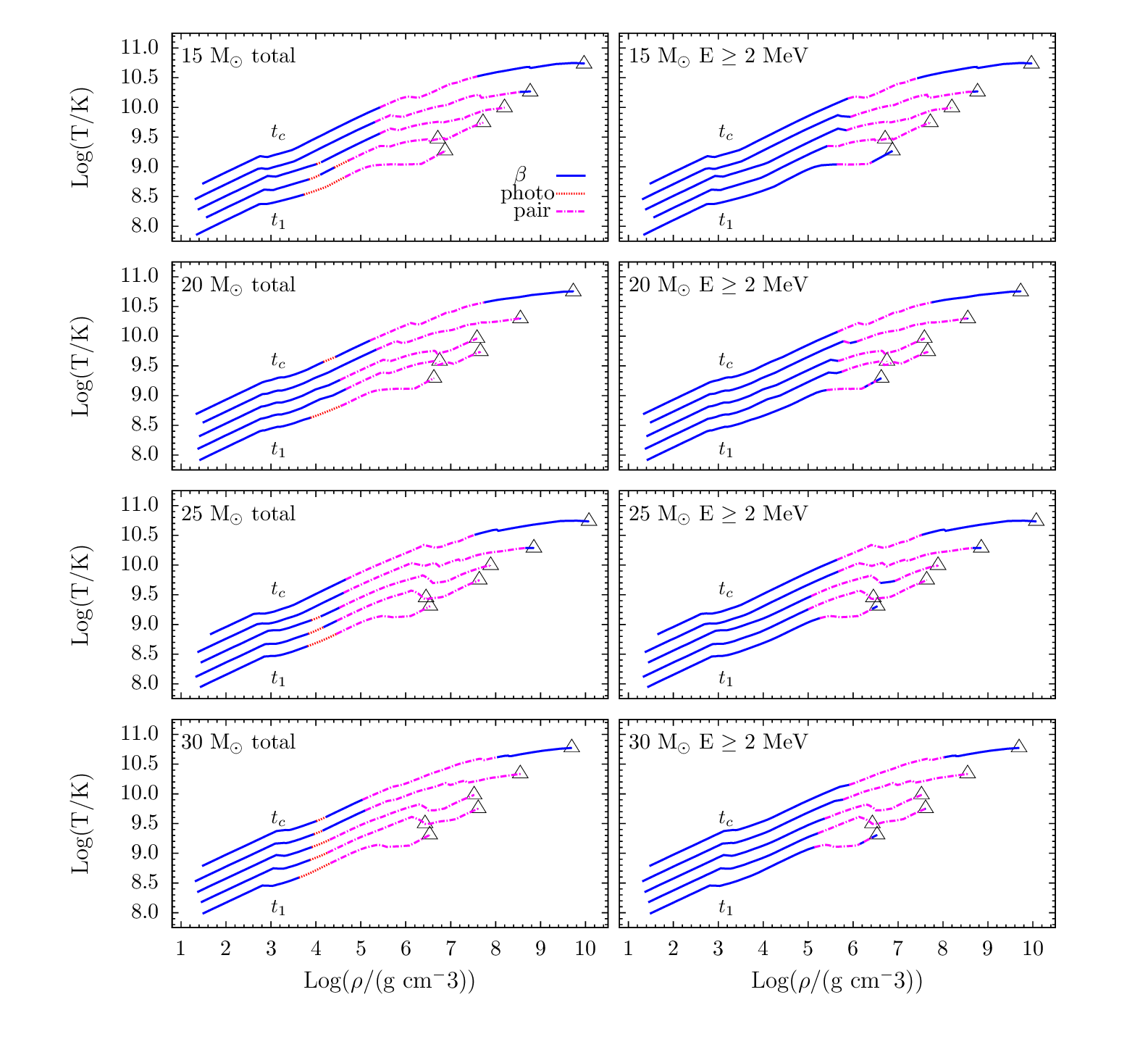}
\caption{{\it Left}:  Origin of the dominant \n\ emissivity as calculated by MESA as a function of both temperature and density, for the same time instants $t_n$ as in fig. \ref{evolution} ($n=1,5$, from bottom to top). Each curve describes the temperature and density encountered at different radii within the star at a given time $t$.  The different dashings/colors indicate which process dominates the total emissivity (see legend).  For better visibility, the curve for the time $t_n$ is shifted upwards vertically by $0.2(n-1)$ units.  The selected points in Table \ref{pointstab} are marked.
{\it Right}:  the same figure, but for the emissivity of potentially detectable \ns\ (with energy $E \geq$ 2 MeV). }
\label{ourEmiss}
\end{figure*}

Fig. \ref{ourEmiss} shows the temperature-density profiles, i.e., the temperature as a function of the density, for each star model at the times $t_n$.  The dashings/colors in the curves indicate which process contributes most strongly to the total and partial emissivities, $Q$ and $\qth$.  As expected from prior literature \citep{Itoh:1996im}, generally $Q$ is mostly due to $\beta$ processes at lower density. Small islands of photoneutrino preponderance in are observed at $\rho \sim 10^4~{\rm g~ cm^{-3}}$, and  pair production dominates for $\rho \gta 10^6~{\rm g ~cm^{-3}}$ and $t \lta t_4$.  At $t=t_5=t_c$, an extended region of $\beta$ dominance appears in the core, at $\rho \gta 10^{7.5}~{\rm g ~cm^{-3}}$.  This phenomenon is consistent with the rapid increase in the rate of neutronization (electron capture) in the core as the onset of collapse approaches.   Indeed, we find that at $t = t_{5} = t_{c}$ electron capture is responsible for $\sim 99\%$ of the total emissivity, $Q$, while the contribution of $\beta$ decay is suppressed (only $\lesssim 1\%$ contribution), as noted above in connection with the electron degeneracy. 

Generally, the partial emissivity, $\qth$, follows the same trends as $Q$, with the major difference that $\beta$ dominance is much more extended to high density, and islands of $\beta$-dominated emissivity are seen for $\rho Y_e$ as high as $\rho Y_e  \sim 10^7~{\rm g~ cm^{-3}}$.  In the core of the star, $\qth$ is always dominated by either pair neutrinos or $\beta$ processes, with the same region of fast neutronization as noted for $Q$. 

Since submission of this work, \citet{2016PhRvD..93l3012Y} has shown the time evolution of the neutrino signal from stars with initial masses of 12, 15, and 20 $M_{\odot}$ models evolved up to core collapse. They find that pair production dominates over $\beta$ processes up to a few hours before collapse, with $\beta$ processes only becoming dominant a few seconds before collapse. Due to this dominance of pair neutrinos, \citet{2016PhRvD..93l3012Y} focus only on that process in their study of spectra and detected events, while $\beta$ interactions are included only in the calculation of total energy loss and abundance evolution of their models. We find $\beta$ decays to be comparable to pair production from $\sim1$ day before core collapse and dominant from $\sim0.5$hr before collapse.  As such, we include detailed calculations of the $\beta$ spectrum as well as thermal processes.  Differences in the neutrino signal can be attributed to differences in the treatment of convection and convective overshooting, leading to differences in the core structure. The core masses in \citet{2016PhRvD..93l3012Y} are larger than ours by $\sim5\%$, an indication of a stronger convective overshoot treatment. For the 15 $M_{\odot}$ model, we found our core temperature and densities to be larger ($\sim 30$\% and $\sim70$\% respectively). These higher temperatures will lead to an increase in production of nuclei undergoing $\beta$ processes, relative to that of \citet{2016PhRvD..93l3012Y}.   

Summarizing, the results of this section  suggest the importance of $\beta$ processes for  the detectable region of the parameter space: late times (where the \n\ luminosity is higher) and the highest energy part of \n\ spectrum. The \n\ energy spectra are the focus of the next section.

\subsection{Spectra}
\label{spectraresults}

\begin{sidewaystable}
\caption{Selected points in the evolution of the star. The first two columns give the instants of time (with $t=0$ the time of collapse, see text) and  the volume-integrated \n\ emissivity at these times.  The following columns refer to specific points inside the star, for which the \n\ spectra were calculated. All points are at the core of the star.  Column 3 specifies the evolutionary stage of the star.  For each set of point and time, columns 5-11 specify the temperature, density, electron fraction, electron degeneracy, radial coordinate, and \n\ emissivity (total of all flavors) $Q$ and $Q_{th}$.  }
\label{pointstab}
\begin{tabular}{cccccccccccc}
\hline
\hline
\multicolumn{9}{c}{15$M_{\odot}$}\\
  Time (hr) &  $Q_{int}$ (MeV/s)  & Stage & Point & $\log{\left(T/K\right)}$  & $\log\left(\rho/\rm g~cm^{-3}\right)$ & $Y_{e}$ & $\eta_{e}$ & $\log(R/{R_\odot})$ & $Q$ (MeV cm$^{-3}$  s$^{-1}$) & $Q_{th}$ (MeV cm$^{-3}$  s$^{-1}$) \\
  \hline
$t_1 = -14425.2$ & 3.499$\times10^{48}$  & Begin core O burning & (c1) & 9.266 & 6.874 & 0.498 & 2.069 & -4.118 & $1.2081\times10^{23}$ & $1.8006\times10^{22}$ \\
\multirow{2}{*}{$t_2 = -24.1$} & \multirow{2}{*}{9.276$\times10^{50}$} & \multirow{2}{*}{Core Si burning} & (s2) & 9.268 & 6.712 & 0.493 & 1.449 & -2.315 & $1.215\times10^{23}$ &  $2.211\times10^{21}$ \\
 & & & (c2)  & 9.544 & 7.718 & 0.471 & 2.895 & -4.560 & $1.347\times10^{26}$ &  $5.398\times10^{25}$ \\
$t_3 = -7.445$ & 3.212$\times10^{51}$  & End core Si burning & (c3) & 9.594 & 8.197 & 0.458 & 4.489 & -4.560 & $2.501\times10^{26}$ & $1.325\times10^{26}$ \\
$t_4 = -0.479$ & 1.708$\times10^{52}$  & 1/2 hour pre-collapse & (c4) & 9.658 & 8.769 & 0.445 & 6.815 & -4.750 & $1.787\times10^{27}$ & $9.577\times10^{26}$ \\
$t_5 = t_c$ & 1.749$\times10^{55}$  & Begin collapse & (c5) & 9.929 & 9.969 & 0.432 & 10.188 & -4.150 & $1.834\times10^{32}$ & $1.696\times10^{32}$ \\
\hline
\hline
\multicolumn{9}{c}{20$M_{\odot}$}\\
  Time (hr) &  $Q_{int}$ (MeV/s)  & Stage & Point & $\log{\left(T/K\right)}$  & $\log\left(\rho/\rm g~cm^{-3}\right)$ & $Y_{e}$ & $\eta_{e}$ & $\log(R/{R_\odot})$ & $Q$ (MeV cm$^{-3}$  s$^{-1}$) & $Q_{th}$ (MeV cm$^{-3}$  s$^{-1}$) \\
  \hline
$t_1 = -2596.2$ & 2.479$\times10^{49}$ & Begin core O burning & (c1)  & 9.292 & 6.626 & 0.498 & 1.009 & -3.994 & $3.006\times10^{23}$ & $3.560\times10^{22}$ \\
\multirow{2}{*}{$t_2 = -24.52$} & \multirow{2}{*}{1.645$\times10^{51}$}  & \multirow{2}{*}{Core Si burning} & (s2) & 9.319 & 6.753 & 0.494 & 1.239 & -2.307 & $5.186\times10^{23}$ &  $1.886\times10^{22}$ \\
&  & & (c2)   & 9.539 & 7.661 & 0.488 & 2.770 & -4.339 & $1.261\times10^{26}$ &  $5.309\times10^{25}$ \\
$t_3 = -11.507$ & 2.647$\times10^{51}$  & End core Si burning & (c3) & 9.561 & 7.586 & 0.481 & 2.256 & -4.314 & $2.376\times10^{26}$ & $1.079\times10^{26}$ \\
$t_4 = -0.465$ & $4.281\times10^{52}$  & 1/2 hour pre-collapse & (c4) & 9.690 & 8.552 & 0.450 & 5.037 & -4.636 & $3.471\times10^{27}$ &  $2.265\times10^{27}$\\
$t_5 = t_c$ & 1.343$\times10^{55}$ & Begin collapse & (c5)  & 9.945 & 9.728 & 0.437 & 7.998 & -4.028 & $6.761\times10^{31}$ & $6.348\times10^{31}$ \\
\hline
\hline
\multicolumn{9}{c}{25$M_{\odot}$}\\
  Time (hr) &  $Q_{int}$ (MeV/s)  & Stage & Point & $\log{\left(T/K\right)}$  & $\log\left(\rho/\rm g~cm^{-3}\right)$ & $Y_{e}$ & $\eta_{e}$ & $\log(R/{R_\odot})$ & $Q$ (MeV cm$^{-3}$  s$^{-1}$) & $Q_{th}$ (MeV cm$^{-3}$  s$^{-1}$) \\
  \hline
$t_1 = -1402.2$ & 6.345$\times10^{49}$ & Begin core O burning & (c1) & 9.306 & 6.537 & 0.498 & 0.652 & -3.932 & $4.543\times10^{23}$ & $5.642\times10^{22}$ \\
\multirow{2}{*}{$t_2 = -23.79$} & \multirow{2}{*}{2.133$\times10^{51}$} & \multirow{2}{*}{Core Si burning} & (c2)  & 9.545 & 7.636 & 0.481 & 0.870 & -4.229 & $4.174\times10^{22}$ &  $3.542\times10^{20}$ \\
 &  & & (s2)  & 9.545 & 7.636 & 0.481 & 2.594 & -4.229 & $1.449\times10^{26}$ &  $5.883\times10^{25}$ \\
$t_3 = -11.81$ & 6.164$\times10^{51}$  & End core Si burning & (c3) & 9.591 & 7.890 & 0.463 & 3.112 & -4.383 & $3.052\times10^{26}$ & $1.418\times10^{26}$ \\
$t_4 = -0.536$ &  $1.695\times10^{52}$ & 1/2 hour pre-collapse & (c4) & 9.684  & 8.851 & 0.443 & 6.886 & -4.804 & $6.527\times10^{27}$ & $4.021\times10^{27}$  \\
$t_5 = t_c$ & 1.921$\times10^{55}$  & Begin collapse & (c5) & 9.934 & 10.078 & 0.428 & 11.046 & -5.112 & $3.227\times10^{32}$ & $2.979\times10^{32}$ \\
\hline
\hline
\multicolumn{9}{c}{30$M_{\odot}$}\\
  Time (hr) &  $Q_{int}$ (MeV/s)  & Stage & Point & $\log{\left(T/K\right)}$  & $\log\left(\rho/\rm g~cm^{-3}\right)$ & $Y_{e}$ & $\eta_{e}$ & $\log(R/{R_\odot})$ & $Q$ (MeV cm$^{-3}$  s$^{-1}$) & $Q_{th}$ (MeV cm$^{-3}$  s$^{-1}$) \\
  \hline
$t_1 = -1063.2$ & 7.100$\times10^{49}$  & Begin core O burning & (c1) & 9.307 & 6.532 & 0.498 & 0.627 & -3.904 & $4.713\times10^{23}$ & $5.552\times10^{22}$ \\
\multirow{2}{*}{$t_2 = -12.04$} & \multirow{2}{*}{5.056$\times10^{51}$}  & \multirow{2}{*}{Core Si burning}  & (s2) & 9.295 & 6.428 & 0.499 & 0.395 & -2.211 & $3.085\times10^{23}$ &  $7.273\times10^{21}$ \\
&  & & (c2)  & 9.550 & 7.612 & 0.490 & 2.481 & -4.264 & $7.867\times10^{25}$ &  $5.705\times10^{25}$ \\
$t_3 = -0.965$ & 9.430$\times10^{51}$  & End core Si burning & (c3) & 9.579 & 7.521 & 0.481 & 1.872 & -4.233 & $4.180\times10^{26}$ & $2.090\times10^{26}$ \\
$t_4 = -0.509$ & 1.156$\times10^{53}$  & 1/2 hour pre-collapse  & (c4) & 9.734 & 8.547 & 0.449 & 4.427 & -4.576 & $9.103\times10^{27}$ & $6.551\times10^{27}$ \\
$t_5 = t_c$ & 1.953$\times10^{55}$  & Begin collapse & (c5) & 9.972 & 9.694 & 0.437 & 7.249 & -4.958 & $1.176\times10^{32}$ & $1.138\times10^{32}$ \\
\hline
\hline
\end{tabular}
\end{sidewaystable}%

We now illustrate the $\nue$ and $\barnue$ energy spectra for selected points inside the star at the times $t_n$. The details about them are given in Table \ref{pointstab}.   These points represent examples of cases when $\beta$ processes contribute substantially to the \n\ spectrum in the detectable energy window.  All points are at the center of the stellar core, except for point (s2), which  is situated at the edge of the core at time $t_{2}$, the beginning of core silicon burning.  In addition to the \n\ emissivities at these specific points,  Table \ref{pointstab} also includes the values of the total emissivity integrated over the entire volume of the star, $Q_{int}$, at the various times.

\begin{figure*}
\includegraphics[width =\linewidth]{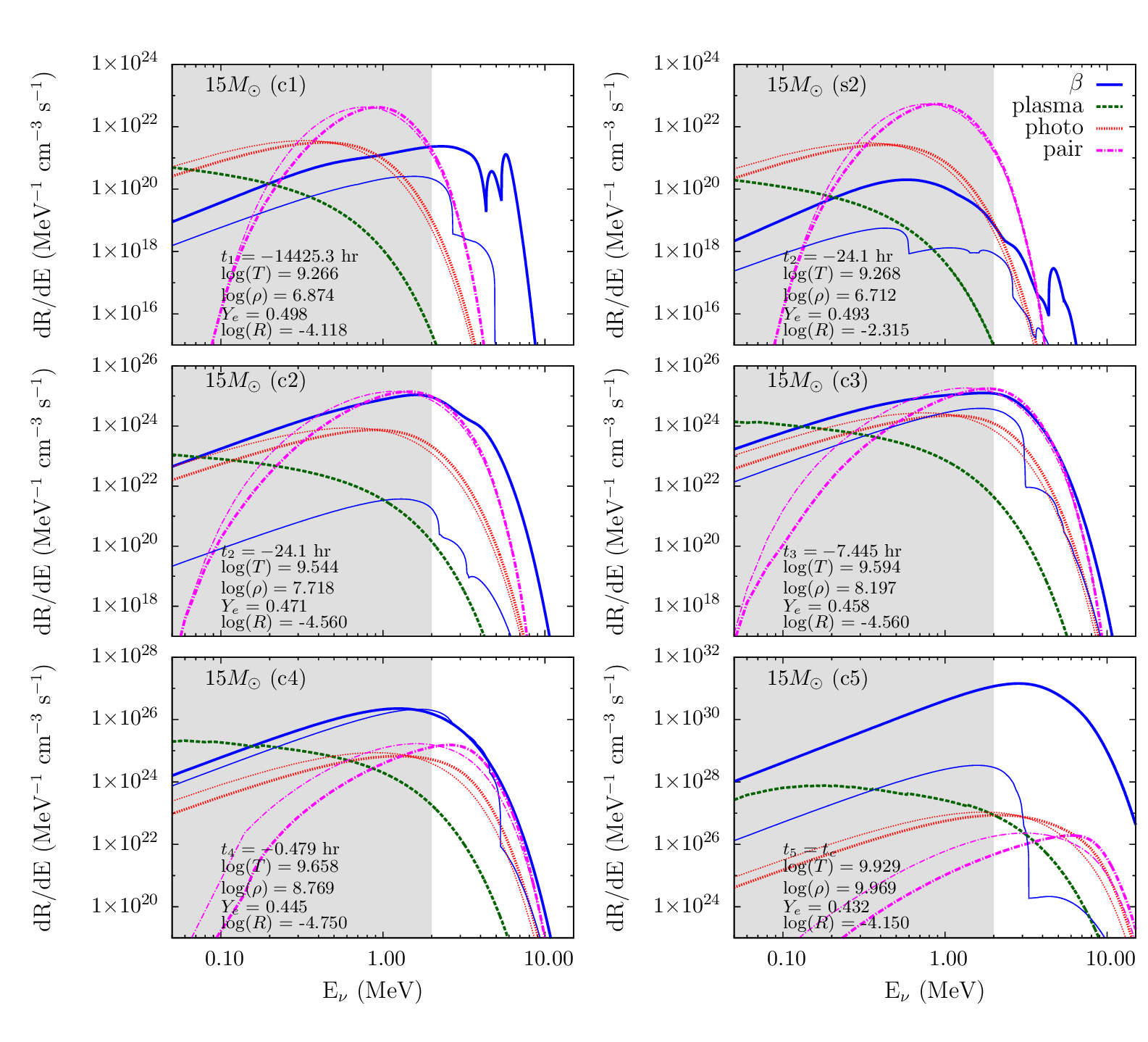}
\caption{Neutrino spectra for different processes, for the 15 $M_{\odot}$ star, for the points (c1)-(c5) and (s2), as described in Table \ref{pointstab}.  Spectra for $\nue$ are shown as thick lines, while $\barnue$ are thin line. The detectable part of the spectrum is shown with light background. Relevant thermodynamic quantities are listed, with units as reported in Table \ref{pointstab}.  
 }
\label{spectra15}
\end{figure*}

\begin{figure*}
\includegraphics[width =\linewidth]{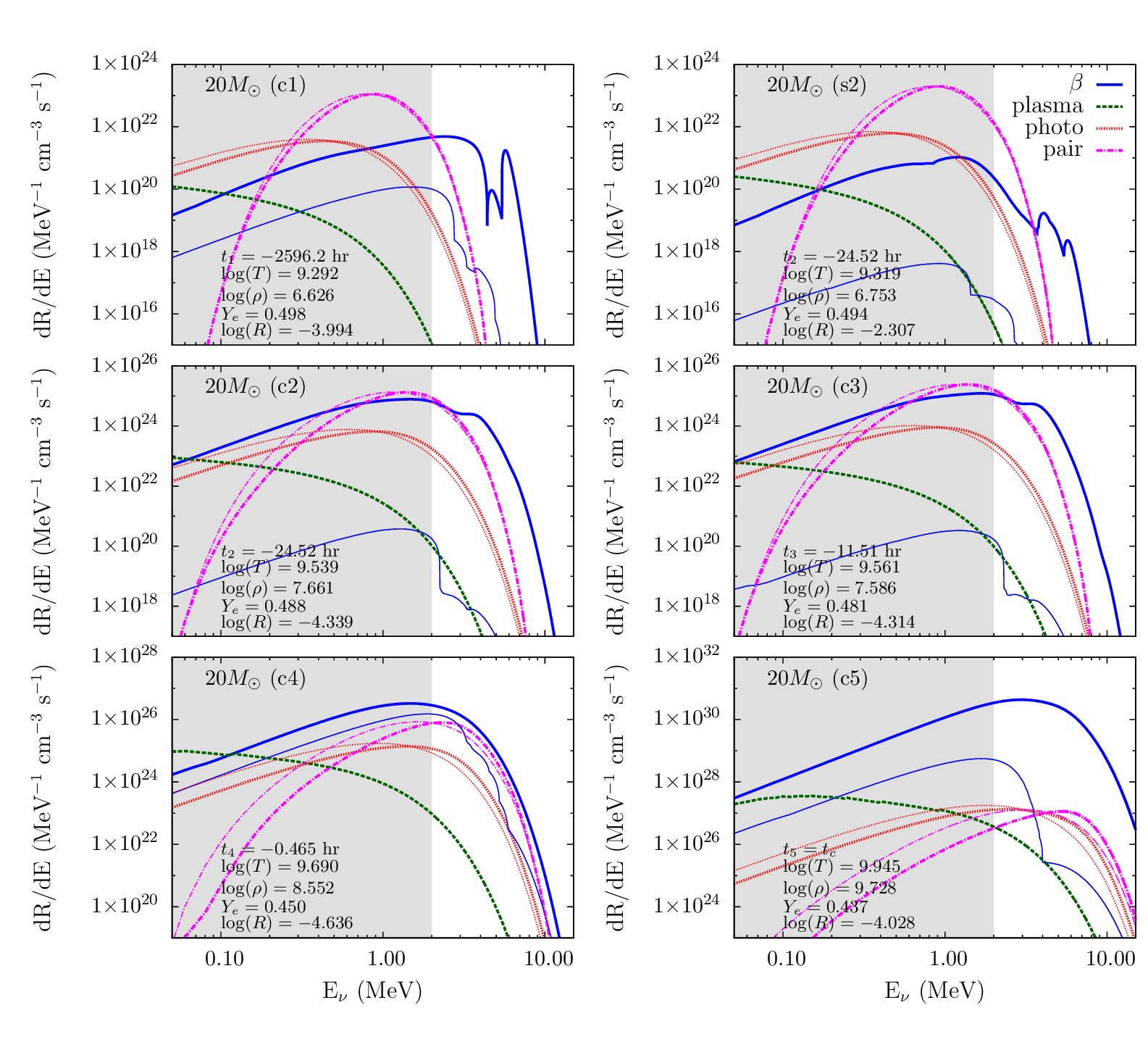}
\caption{The same as fig. \ref{spectra15},  for the points (c1)-(c5) and (s2), of the 20 $M_{\odot}$ star as given in Table \ref{pointstab}.  } 
\label{spectra20}
\end{figure*}

\begin{figure*}
\includegraphics[width =\linewidth]{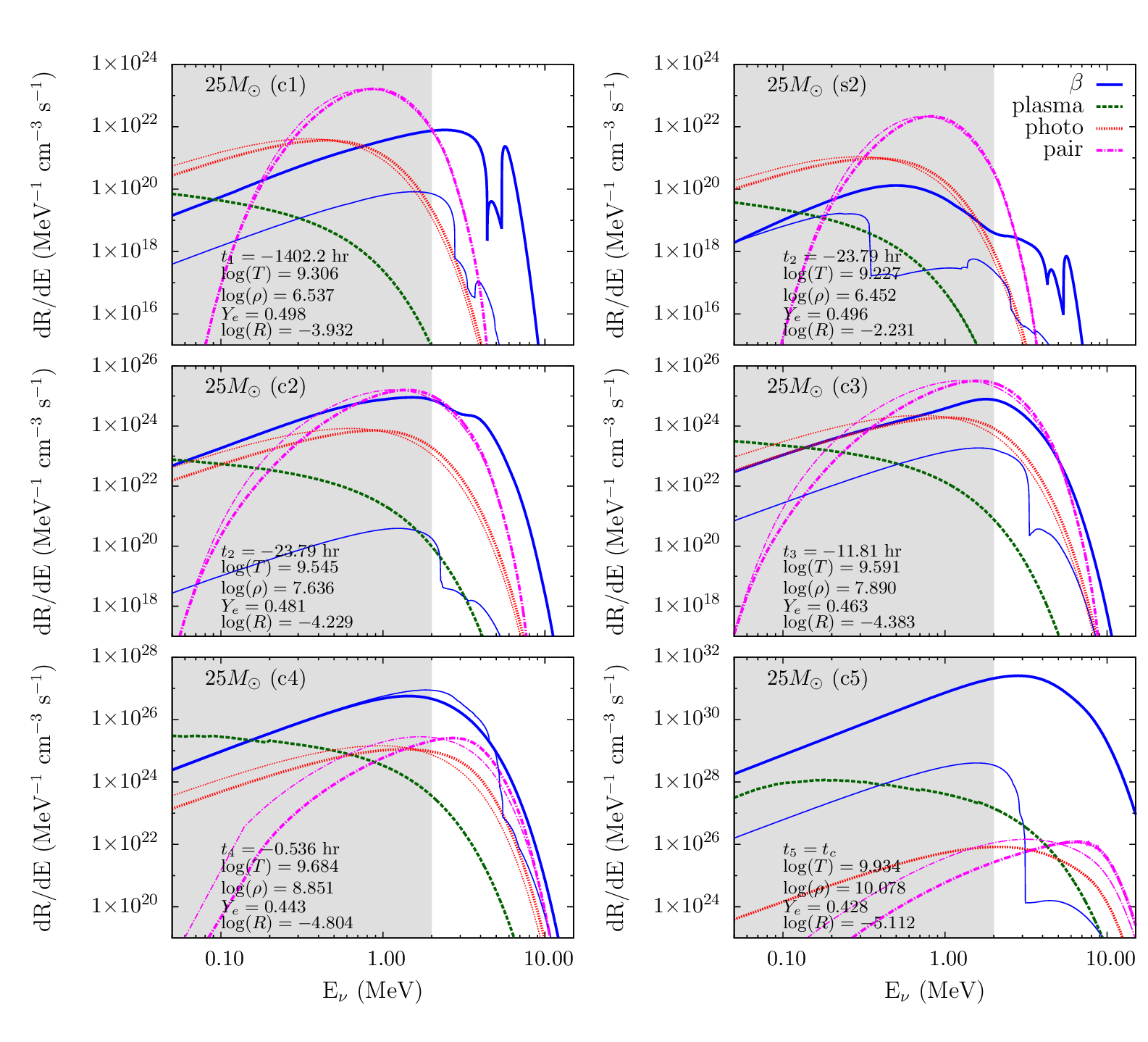}
\caption{The same as fig. \ref{spectra15}, for the points (c1)-(c5) and (s2), of the 25 $M_{\odot}$ star as given in Table \ref{pointstab}.  
 }
\label{spectra25}
\end{figure*}

\begin{figure*}
\includegraphics[width =\linewidth]{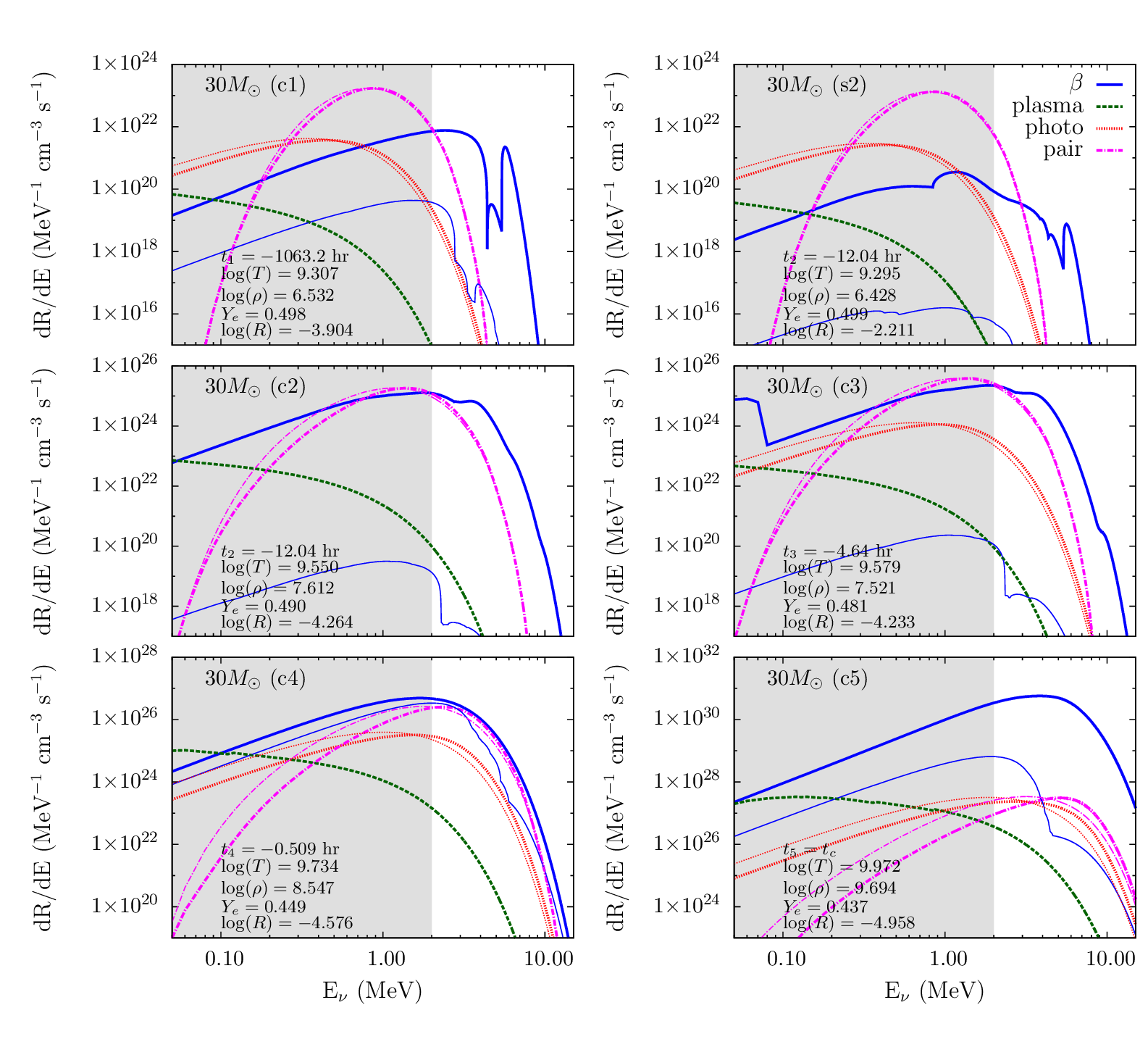}
\caption{The same as fig. \ref{spectra15}, for the points (c1)-(c5) and (s2), of the 30 $M_{\odot}$ star as given in Table \ref{pointstab}.  } 
\label{spectra30}
\end{figure*}

Figs. \ref{spectra15}, \ref{spectra20}, \ref{spectra25} and \ref{spectra30} show the contribution of the different processes to the \n\ spectra for each progenitor model. We see that -- although with the individual differences discussed in sec. \ref{history} -- the main features of the spectra are common to all the stellar models. Specifically, at points (c1) and (s2) pair production has the dominant contribution  at $E \sim $MeV, with plasma \ns\ and photoneutrinos becoming increasingly important, or even dominant, at lower energies.  In the tail of the spectrum, $E \gta E_{th}=2$ MeV, the major contributions to the energy spectrum are from pair production and $\beta$ processes. The latter can dominate by several orders of magnitude at $E \sim 7-10$ MeV, which is a realistic energy threshold for a Mt-mass water Cherenkov detector \citep{Abe:2011ts}.   In the core, the contribution of $\beta$ processes to the spectrum is substantial at all energies for $t\gta t_2$ (points (c2)-(c5)), and dominates by more than one order of magnitude  at the onset of collapse (point (c5)).  Generally, $\beta$ processes contribute more strongly to the $\nue$ spectrum, due to the high rate of electron capture, but still they can play a major role for the $\barnue$ spectrum as well (for points (c4) and (c5), see e.g., bottom of fig. \ref{spectra15} - \ref{spectra30}). 

The structure of the $\beta$ spectra in Figs. \ref{spectra15}-\ref{spectra30} is as expected: one can identify the characteristic smooth shape of $\beta$ decay spectra and, especially at lower temperatures (e.g., point (c1)), the peaks due to electron capture.  At higher temperatures, these peaks are widened by thermal effects (see e.g., \citep{Odrzywolek:2009yl}) and ultimately form a continuum with one another and with the electron capture spectrum (points (c3), (c4) and (c5)). 

We have identified the specific decays that produce the most prominent $\beta$  peaks.  For all progenitor models, at points (c1) and (s2) , where the temperatures and densities are relatively low, the major contributors are isotopes around $A\approx 30$.  In particular, the two peaks at high energy in the $\nue$ spectrum for the point (c1) in all stellar masses are produced by electron capture on $^{31}$S and $^{30}$P.  The high energy peaks in the spectrum for point (s2) are due to electron capture on $^{30}P$ for $15 M_{\odot}$; $^{31}$P, $^{31}$Si and $^{48}$V for $20 M_{\odot}$; $^{31}$Si and $^{31}$S for $25 M_{\odot}$; and $^{31}$S for $30 M_{\odot}$.   The peaks in the $\barnue$ spectra are also due to sulfur, phosphorous and silicon isotopes around $A\approx 30$.  As the temperature and density increases, the isotopes dominating the $\beta$ process spectrum move to higher $A$.  The $\nue$ and $\barnue$ spectra for points (c3) - (c5) have the highest contributions from iron, cobalt, manganese and chromium isotopes, as well as capture reactions on neutrons and protons.  This is consistent with the findings of \citet{Odrzywolek:2009yl}.  The possibility that these decays might, at least in principle, be observed in a \n\ spectrum could serve as motivation for further theoretical study.

\section{Summary and discussion}
\label{sec:discussion}

We have performed a new study of the \n\ emission from a star in the \ps\  phase. This work is the first to combine all the relevant microphysics -- including $\beta$ processes -- with a state-of-the art numerical simulation of stellar evolution.  We were able to obtain, for the first time, accurate and consistent \n\ fluxes and spectra from $\beta$ processes, using the detailed isotopic composition calculated by the MESA code, with a nuclear network of up to 204 isotopes.  The $\nue$ and $\barnue$ emissivities and spectra were calculated for selected times and locations inside the star, and for four stellar progenitors of different masses.  Particular emphasis was placed on the detectable part of the spectrum, above an indicative threshold of 2 MeV. 

It was found that, in part of the parameter space, $\beta$ processes contribute substantially to the detectable \n\ flux, even when they are subdominant to the entire \n\ emissivity (integrated over the entire spectrum). In the last minutes before collapse, the $\nue$ flux from electron capture largely dominates -- by more than one order of magnitude --  the \n\ emission in the core of the star at energies relevant for detection.  Some of the $\beta$ decays that contribute the most, due to having high Q-value, were identified; they would be an interesting target of further study to obtain more reliable spectra above realistic detection thresholds. 

Results for different stellar models show that the time evolution of the \n\ flux is strongly progenitor-dependent, reflecting the faster evolution of more massive stars through the different stages of nuclear burning (see e.g., \citep{Paxton:2010ig}).  Therefore the time distribution of a \ps\ \n\ signal in a detector might be a new tool to learn about \sn\ progenitors. This could be especially interesting in the context of failed \sne: the observation of \ns\ from advanced nuclear burning, combined to an anomalously short \n\ burst (truncated by the direct collapse into a black hole, see e.g., \cite{Sumiyoshi:2007pp}) would unambiguously identify a failed \sn, and help to constrain its progenitor mass. Such constraint would contribute to the debate on what characteristics of the progenitor star (mass, compactness, etc.) are most strongly correlated to direct black hole formation (see e.g., \cite{OConnor:2010moj,Pejcha:2014wda}). 

We find that the $\nue$ and $\barnue$ fluxes should dominate over the $\nux$ ones.  This could, in principle, make \ps\ \ns\ a tool to test \n\ oscillations  by looking for the permutation of energy spectra of the different flavors.  Interestingly, the oscillation pattern might be different from that expected for the post-collapse flux. Specifically, the \ps\ flux might be free from the still poorly understood collective oscillation effects --  driven by \n-\n\ coherent scattering -- (see e.g., \citep{0904.0974,arXiv:1001.2799})  that are active when the number flux of \ns\ is high and the matter density profile is suppressed behind the launched shockwave \citep{Hannestad:2006nj,Mirizzi:2015eza}. Without the complication of collective effects, oscillations of \ps\ \ns\ might offer a particularly clean test of the matter-driven flavor conversion (MSW effect \citep{Wolfenstein:1977ue,Mikheev:1986wj,Mikheev:1986if}). 

A step forward towards a study of the detectability of \ps\ \ns\ will be to integrate the emissivity over the whole star's core and inner shells and for several time steps, so as to obtain the total \n\ flux, its spectrum,  and its time evolution.  This will be the subject of our future work.

\begin{acknowledgments}
We are deeply grateful to F.~X.~Timmes for very insightful comments and encouragement, and thank A. Odrzywolek for fruitful discussion. We are also indebted to the anonymous referee for stimulating feedback.   K.~M.~Patton and C.~Lunardini acknowledge the National Science Foundation grant number PHY-1205745 and the Department of Energy award DE-SC0015406.  K. M. Patton also acknowledges the US Dept of Energy grant DE-FG02-00ER41132. R.~Farmer acknowledges support from NASA under the Theoretical and Computational Astrophysics Networks (TCAN) grant
NNX14AB53G and NSF under the Software Infrastructure for.Sustained Innovation (SI$^2$) grant 1339600.

\end{acknowledgments}

\bibliography{bibUsed}

\end{document}